\documentclass[aps,prl,floatfix,twocolumn,footinbib,showpacs]{revtex4}
\usepackage{amsmath,amssymb}
\usepackage{graphicx,color}
\usepackage{subfigure}
\usepackage{bm}
\begin{document}

\title{Electron spin relaxation in graphene: the role of the substrate}

\author{Christian Ertler}
\email[]{christian.ertler@physik.uni-regensburg.de}
\author{Sergej Konschuh}
\author{Martin Gmitra}
\author{Jaroslav Fabian}
\affiliation{Institute for Theoretical Physics, University of
Regensburg, 93040 Regensburg, Germany}

\begin{abstract}

Theory of the electron spin relaxation in
graphene on the SiO$_2$ substrate is developed. Charged impurities and polar
optical surface phonons in the substrate induce
an effective random Bychkov-Rashba-like spin-orbit coupling field which
leads to spin relaxation by the D'yakonov-Perel' mechanism.
Analytical estimates and Monte Carlo simulations show that the corresponding spin
relaxation times are between micro- to milliseconds, being only weakly
temperature dependent. It is also argued that the presence of adatoms on graphene
can lead to spin lifetimes shorter than nanoseconds.

\end{abstract}


\maketitle

Since the experimental realization of graphene, a single stable
2D-monolayer of carbon atoms arranged in a honeycomb lattice,
considerable research has been done to enlighten its peculiar
electronic transport properties originating from the Dirac-like band
structure at the $K$ and $K'$ points in the momentum space
\cite{Geim2007:NatureMat}. Long spin relaxation times and phase
coherence lengths in graphene are expected based on the weak atomic
spin-orbit coupling in carbon ($Z=6$). However, recent spin
injection measurements based on a non-local spin valve
geometry \cite{Tombros2007:Nature, Tombros2008:PRL,
Jozsa2008:PRL} revealed surprisingly short spin relaxation times of
only about 100-200 ps, being only weakly dependent on the charge
density and temperature. These results appear puzzling,
although the low mobilities of the samples (about 2000 cm$^2$/Vs)
suggest that the measured spin relaxation times are likely
due to extrinsic effects \cite{Tombros2007:Nature}.

Very recent experiments on the charge transport in graphene affirmed
the importance of the underlying substrate
\cite{Chen2008:NatureP,Adam2008:SSC, Sabio2008:PRB}. At low
temperatures the transport properties have been shown to be
dominated by scattering from the charged impurities residing in the
substrate \cite{Adam2007:PNAS, Adam2008:SSC}. The conductivity of
graphene placed on a SiO$_2$ substrate starts to decrease above 200
K. The observed temperature and density dependence of the
resistivity are most likely explained by remote phonon scattering
due to occurrence of polar optical surface modes in the substrate
\cite{Fratini2008:PRB,Guinea2008:JLTP, Chen2008:NatureN}.

These findings naturally raise the question if (i) charged
impurities and (ii) remote surface phonons are also relevant for the
spin relaxation in graphene. As argued here both mechanisms provide
a temperature-dependent, random spin-orbit coupling field, which limits the
spin relaxation via the D'yakonov-Perel' (DP) mechanism
\cite{Dyakonov1972:SPSS, Dyakonov:1984b, Fabian2007:APS}.
The calculated spin relaxation times are micro to milliseconds.
In addition, we give estimates for the
spin relaxation times due to the possible presence of adatoms on graphene.
For reasonable adatom densities the spin lifetimes can be lower
than nanoseconds.

Several other mechanisms have already been
investigated theoretically, such as the spin relaxation due to the
corrugations (ripples) of graphene and due to exchange interaction
with local magnetic moments \cite{Huertas2007:EPJ}, or spin-orbit coupling
mediated relaxation based on boundary scattering, heavy impurities,
and effective gauge fields due to topological disorder
\cite{Huertas2008:condmat}.


\begin{figure}
\centering
\subfigure{\includegraphics[width=0.47\linewidth]{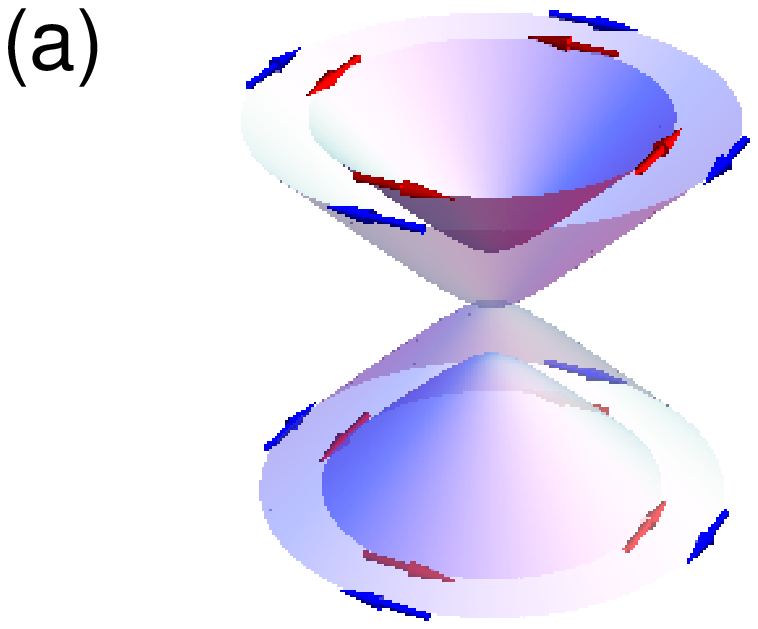}}
\subfigure{\includegraphics[width=0.35\linewidth]{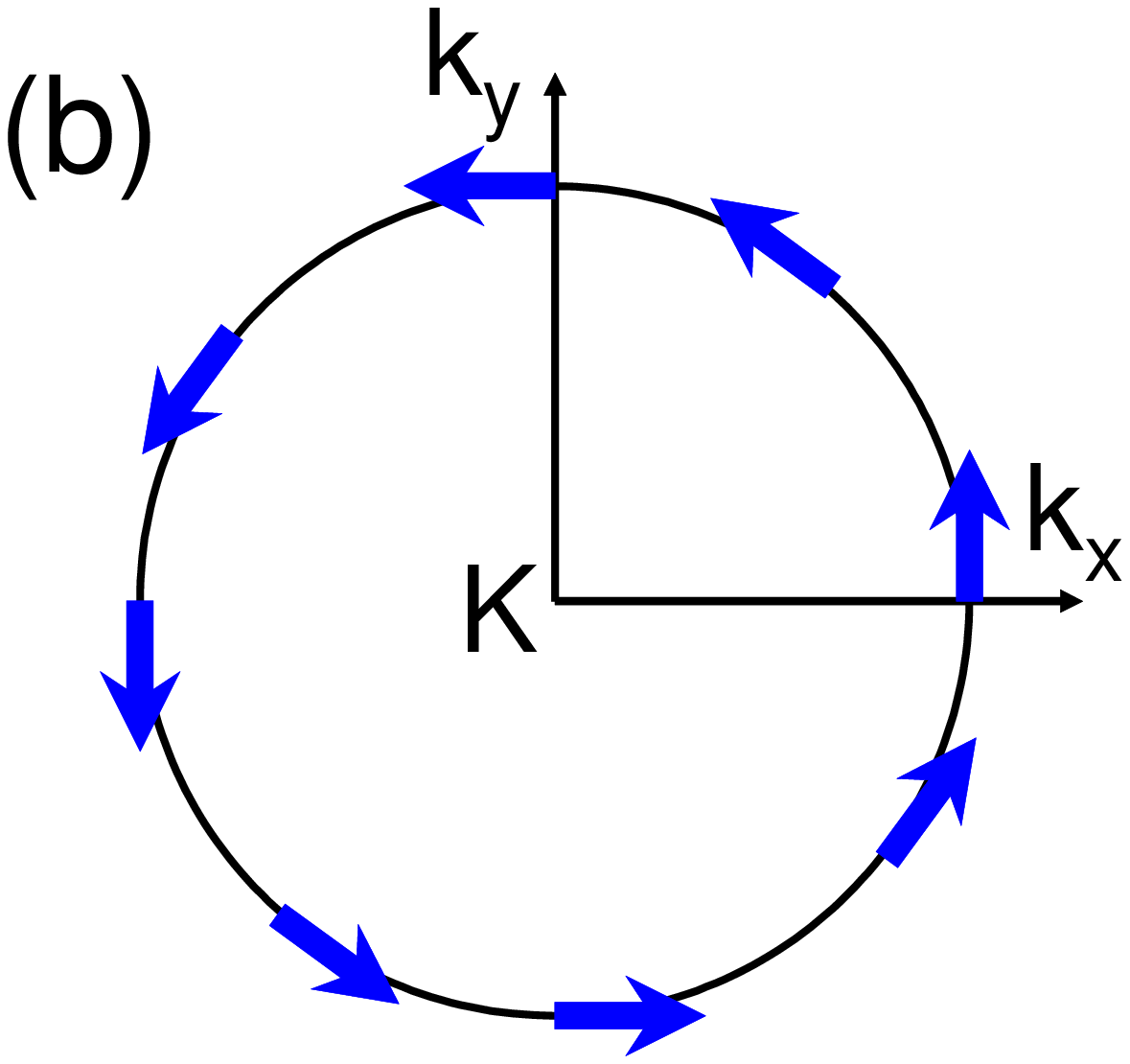}}\\
\subfigure{\includegraphics[width=0.45\linewidth]{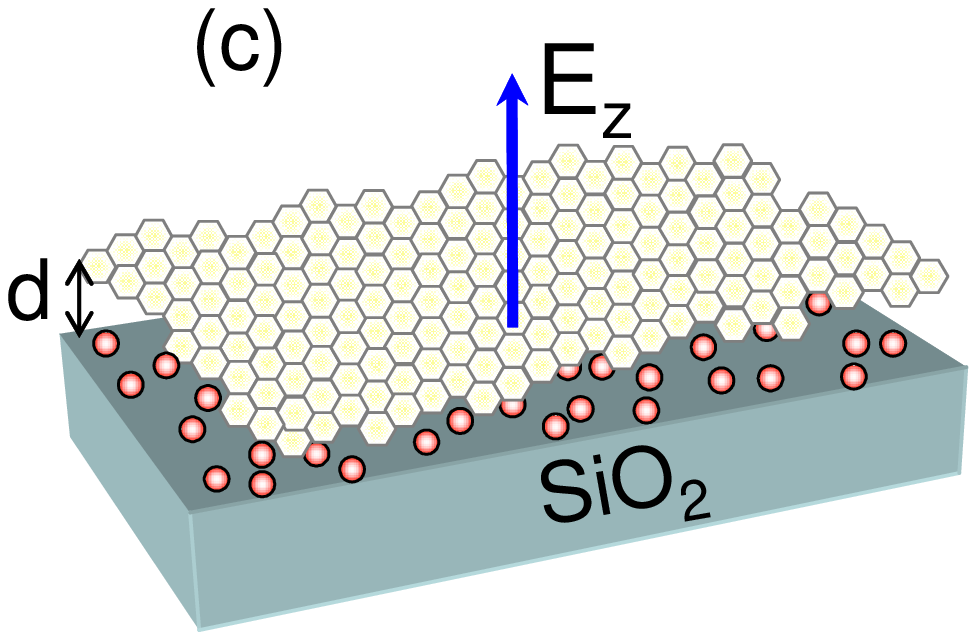}}
\subfigure{\includegraphics[width=0.49\linewidth]{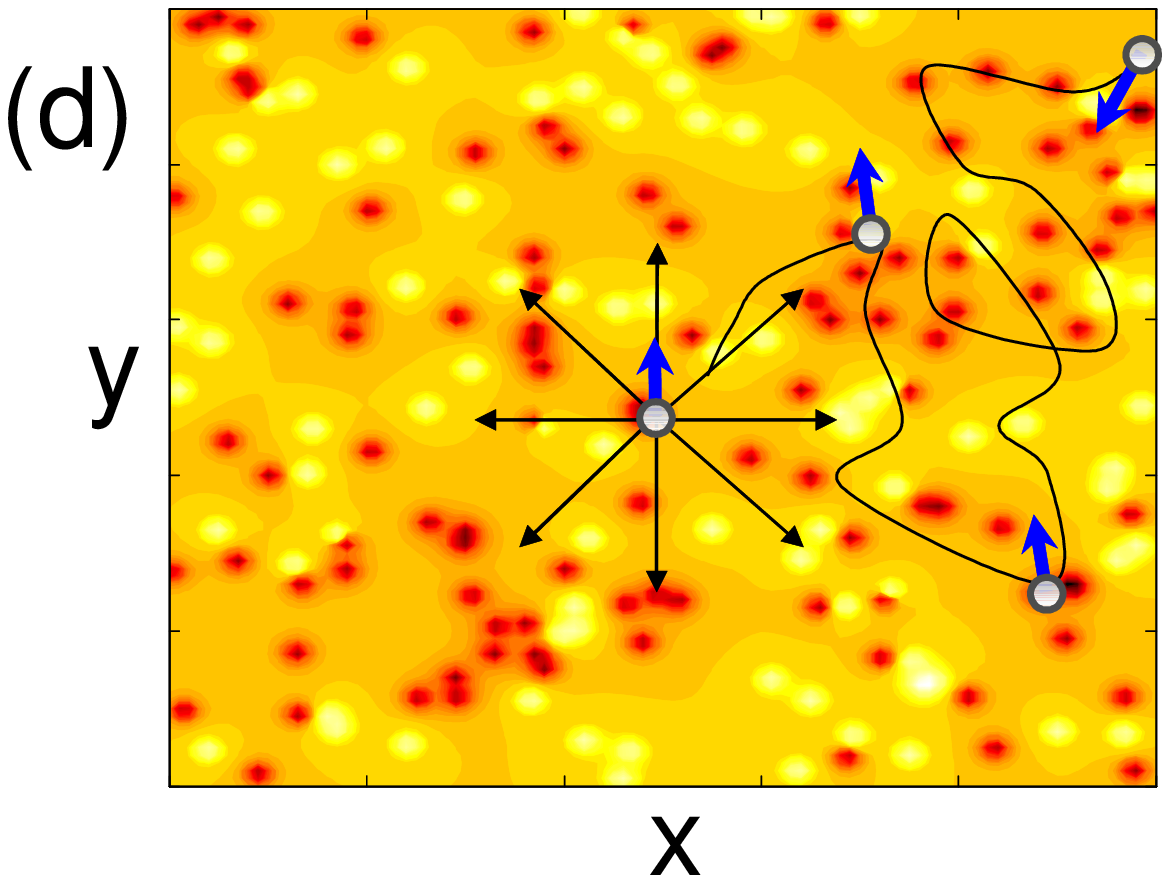}}\\
\caption{\label{fig:scheme} (Color online) (a) The Dirac cones when spin-orbit
coupling is included. The arrows indicate the spin vectors
$\mathbf{n}^K_{m\nu}$ as defined in the text. (b) Effective
magnetic field directions (Bychkov-Rashba field) along the Fermi
circle of electrons at the $K$-point (the field is the same at the
$K'$-point). (c) Graphene layer on the top of a SiO$_2$ substrate
with charged impurities which induce an electric field component
$E_z$ perpendicular to the plane breaking the inversion symmetry of
graphene. (d) Illustration of the spin relaxation in a spatially
random potential due to the charged carriers. In the Monte-Carlo
simulations the spin dynamics is sampled over random trajectories
with different initial momenta.}
\end{figure}

Near the $K$ and $K'$ points the carrier dynamics can be described
 by an effective low energy
Hamiltonian \cite{DiVincenzo1984:PRB} of the form $H_0 = \hbar
v_\mathrm{f}(\tau_z\sigma_x k_x+\sigma_y k_y)$. Here, $v_f = 10^6 $
m/s denotes the Fermi velocity, $\mathbf{k}$ is the wave vector with
respect to $K(K')$, and $\mathbf{\tau}$ and $\mathbf{\sigma}$ are
the Pauli matrices with $\tau_z = \pm 1$ describing the states at
the $K$ and $K'$-points, respectively, and $\sigma_z =\pm 1$
describing the states on the A and B sublattice of the honeycomb
lattice. The inclusion of the microscopic spin-orbit (SO)
interaction results in an additional term in the effective
low-energy Hamiltonian, $H_I = -\lambda_I+\lambda_I\tau_z\sigma_z
s_z$ as shown either by group theoretical arguments
\cite{Kane2005:PRL} or by second order perturbation theory of a
microscopic tight-binding model \cite{Min2006:PRB, Yao2007:PRB,
Huertas2006:PRB}. Here, the real spin is represented by the $s_z$
Pauli matrix and  $\lambda_I$ denotes the intrinsic SO-constant. The
intrinsic SOC opens a gap $\Delta_I = 2\lambda_I$ at the Dirac
point, making graphene theoretically a spin Hall insulator
\cite{Kane2005:PRL}. Recent first-principles calculations give
$\Delta_I = 0.024$ meV \cite{Gmitra2009:arXiv}, large enough to
influence electronics of intrinsic or weakly charged graphene only
somewhat below 1~K.

If an electric field is applied perpendicular to the graphene plane,
the inversion symmetry is lifted and group theory allows for an
additional Bychkov-Rashba (BR) term of the form $H_{\mathrm{BR}} =
\lambda_{\mathrm{BR}}(\tau_z\sigma_x s_y-\sigma_y s_x)$
\cite{Kane2005:PRL, Min2006:PRB,Huertas2006:PRB, Rashba2009:PRB}.
From first-principles calculations \cite{Gmitra2009:arXiv} a linear
relationship between the BR-constant and the electric field,
$\lambda_{\mathrm{BR}}(\mathbf{r}) = \zeta_{\mathrm{BR}}
E_z(\mathbf{r})$, is found, with $\zeta_{\mathrm{BR}} = 0.005$
meV/(V/nm). The proper knowledge of $\lambda_{\mathrm{BR}}$ is of
great importance for our quantitative results below, since in the
DP-mechanism the spin relaxation rate depends quadratically on
$\lambda_{\mathrm{BR}}$.

The resulting effective $8\times8$ Hamiltonian $H_\mathrm{eff} = H_0
+ H_I + H_{\mathrm{BR}}$ is easily diagonalized yielding the same
eigenvalues at the $K$ and $K'$ points,
\begin{equation}
\varepsilon_{m \nu} = \nu(\lambda_{\mathrm{BR}}-\nu\lambda_I)+
m\sqrt{\varepsilon^2+(\lambda_{\mathrm{BR}}-\nu\lambda_I)^2}
\end{equation}
with $\varepsilon = \hbar v_f |k|$, $\nu= \pm 1$, and the band index
$m=1$ for electrons $(e)$ and $m=-1$ for holes $(h)$, respectively.
We define spin vectors $\mathbf{n}_{m\nu}^\tau =
 \mathbf{s}_{m\nu}^\tau/| \mathbf{s}_{m\nu}^\tau|$
as normalized expectation values of the spin operator
$\mathbf{s}_{m\nu}^\tau=\langle\psi|\hat{\mathbf{S}}|\psi\rangle$
with respect to the eigenstates $|\psi\rangle =
|\tau,\mathbf{k},m,\nu\rangle$ $(\tau = K,K')$ of the total
Hamiltonian $ H_\mathrm{eff}$. The vectors are in-plane and result
in $\mathbf{n}^\tau_{e+} =  \mathbf{n}^\tau_{h-} =
(-\sin\varphi,\cos\varphi,0)$ and $\mathbf{n}^\tau_{e-} =
\mathbf{n}^\tau_{h+} = (\sin\varphi,-\cos\varphi,0)$ with $\varphi$
denoting the polar angle of the wave vector $\mathbf{k}$. In the
case of $\varepsilon \gg \lambda_R+\lambda_I$, i.e., if the Fermi
energy is much greater than $\approx 0.03$ meV (a condition usually
fulfilled in gated or doped graphene) the electron and hole motion
can be decoupled. By successive unitary rotation of $H_\mathrm{eff}$
first into the eigenbasis of $H_0$ and then into the spin basis with
respect to the direction $\mathbf{n} = (-\sin\varphi,\cos\varphi,0)$
an effective BR-type $2\times 2$ Hamiltonian can be obtained for
both holes and electrons,
\begin{equation}
\tilde{H}_\mathrm{eff} =
m(\varepsilon-\lambda_I)+m\lambda_{\mathrm{BR}}
\mathbf{n}(\mathbf{k})\cdot\mathbf{s}
\end{equation}
with $\mathbf{s}$ denoting the Pauli spin matrices.
$\tilde{H}_\mathrm{eff}$ is the same for $K$ and $K'$, as guaranteed
by time reversal symmetry.

Comparison with the original BR-Hamiltonian in semiconductor
heterostructures of the form $H_{\mathbf{k}} =
\hbar\mathbf{\Omega}(\mathbf{k})\cdot\mathbf{s}/2$ shows that SOC
coupling in graphene effectively acts on the electrons spin as a
in-plane magnetic field of {\em constant} amplitude but
$\mathbf{k}$-dependent direction, as illustrated in
Fig.~\ref{fig:scheme}(a) and (b). In this effective field the spin
precesses with a frequency of $\Omega =
2\lambda_{\mathrm{BR}}/\hbar$. As shown by D'yakonov and Perel'
\cite{Dyakonov1972:SPSS, Dyakonov:1984b} random scattering induces
motional narrowing of this spin precession causing spin relaxation.
The spin relaxation rates for the DP-mechanism for the $\alpha$-th
spin component generally result in $1/\tau_{s,\alpha}=\tau^*(\langle
\mathbf{\Omega}_\mathbf{k}^2\rangle -
\langle\mathbf{\Omega}_\alpha^2\rangle)$ with $\tau^*$ denoting the
correlation time of the random spin-orbit field and
$\langle\ldots\rangle$ indicates averaging over the Fermi surface.
Due to the polar angle dependence, in graphene the correlation time
exactly coincides with the momentum relaxation time $\tau^* =
\tau_p$ \cite{Dyakonov:1984b, Fabian2007:APS}. Hence, for graphene the spin
relaxation time results in $1/\tau_{s,z} = \tau_p (2
\lambda_{\mathrm{BR}}/\hbar)^2$ and $\tau_{s,\{x,y\}}= 2\tau_{s,z}$.

\begin{figure}
\includegraphics[width=0.9\linewidth]{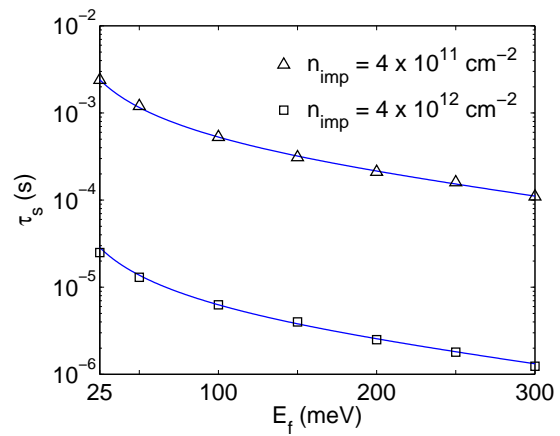}
\caption{\label{fig:tausimp}(Color online) Calculated spin relaxation time
$\tau_s$ as a function of the Fermi energy $E_f$, taking into
account only impurity scattering, for two different impurity
densities in the substrate at $T = 0$ K. The symbols indicate
MC-simulation results and the solid lines are analytic fits of the
form $1/\tau_s = \tau_{\mathrm{imp}}(E_f) \Omega_\mathrm{eff}^2$
with $\Omega_\mathrm{eff} = 3.3\times10^9$ s$^{-1}$ (for squares)
and $\Omega_\mathrm{eff} = 1.1\times10^8$ s$^{-1}$ (for triangles).}
\end{figure}


First, we investigate spin relaxation due to charged impurities
residing in the substrate as schematically illustrated in
Fig.~\ref{fig:scheme}(c). Impurity scattering is a dominant
scattering mechanism governing the transport properties of graphene
\cite{Adam2007:PNAS, Chen2008:NatureP, Adam2008:SSC}. Due to the
fluctuations of the impurity concentration a random unscreened
electric field perpendicular to the graphene plane, and hence a
spatially random BR-field $\lambda_{\mathrm{BR}}(\mathbf{r}) =
\zeta_{\mathrm{BR}} E_z(\mathbf{r})$ arises. As shown by Sherman
\cite{Sherman2003:APL} in the case of semiconductor quantum wells,
the randomness of the BR-field in the real space already causes spin
relaxation even without any scattering in the $\mathbf{k}$-space.
The correlation length of the random BR-field is on the scale of the
distance $d$ of the impurity layer from the graphene sheet
\cite{Sherman2003:APL}. Therefore, the spin relaxation time for a
ballistically moving electron can be estimated as follows: if the
electron passes through a domain of the lateral size of the
correlation length of the BR-field, the spin precesses by $\delta
\varphi = \Omega_{\mathrm{BR}} d/v_f$. At some time $t$ the electron
has passed through $t/(d/v_f)$ different domains and in the picture
of a random walk it follows that $\langle \Delta \varphi\rangle =
\delta\varphi\sqrt{t/(d/v_f)}$. The spin is relaxed if
$\langle(\Delta\varphi)^2\rangle \approx1$ yielding the condition
$1/\tau_s \approx 4/\hbar^2 \langle \lambda_{\mathrm{BR}}^ 2\rangle
d/v_f.$ Hence, in a semiclassical picture for the orbital motion
$\mathbf{r}(t)$ of the electron, the spin experiences a random
BR-field {\em both} in the real space (Sherman mechanism) and in the
reciprocal space due to momentum scattering (DP-mechanism).

We numerically calculate the spin relaxation time by performing
Monte Carlo (MC)-simulations for the spin dynamics. For this purpose
we use a random but quenched impurity distribution of a given
density and sample over the random particles trajectories starting
with different initial momenta, as illustrated in
Fig.~\ref{fig:scheme}(d). The trajectories are generated according
to the scattering probability of the screened impurity potentials in
the graphene sheet calculated in the random phase approximation
following Ref. \onlinecite{Adam2008:SSC}.

Along any given
semiclassical trajectory $[\mathbf{r}(t),\mathbf{k}(t)]$ the spin
dynamics can be described by the Bloch equation
\begin{equation}
\frac{d \mathbf{s}}{d t} = \Omega_{\mathrm{BR}}[\mathbf{r}(t)]
(\mathbf{n}[\mathbf{k}(t)]\times \mathbf{s}).
\end{equation}
The spin relaxation time is then calculated by averaging over the
asymptotics of all trajectories, since for times much greater than
the mean free time $t\gg \tau_{\mathrm{mfp}}$ the spin components
relax as $s_\alpha(t) = s_\alpha(0)\exp(-t/\tau_{s,\alpha})$
\cite{Sherman2003:APL}.

Figure \ref{fig:tausimp} shows the calculated spin relaxation time
as a function of the Fermi energy $E_f$ for a dirty SiO$_2$
substrate,  $ n_\mathrm{imp} = 4\times10^{12}$ cm$^{-2}$, and for a
cleaner sample, $ n_\mathrm{imp}=4\times10^{11}$ cm$^{-2}$, taking
into account only impurity scattering. For all simulations we use
the \emph{ab-initio} BR-parameter $\zeta_{\mathrm{BR}} = 0.005$
meV/(V/nm) and an effective impurity distance of $d = 0.4$ nm from
the graphene layer \cite{Adam2008:SSC, Fratini2008:PRB}. The symbols
refer to the MC-simulation results and the solid lines indicate
analytic fits of the form $1/\tau_s = \tau_{\mathrm{imp}}(E_f)
\Omega_\mathrm{eff}^2$ with $\tau_\mathrm{imp}$ denoting the
momentum relaxation time due to impurities. Since the cross section
of the screened long-ranged Coulomb potential is proportional to the
Fermi wavelength $\lambda_f\sim k_f^{-1}$ \cite{Guinea2008:JLTP},
the momentum relaxation time increases with increasing Fermi energy
yielding a decreasing spin relaxation time, as illustrated in
Fig.~\ref{fig:tausimp}


The second important spin relaxation mechanism induced by the
SiO$_2$ substrate is due to polar optical surface phonons. In the
case of SiO$_2$ there are two dominant surface phonons with energies
of $\hbar \omega_s^ {(1)} = 59 $ meV and $\hbar \omega_s^ {(2)} =
155 $ meV, respectively, which provide a temperature dependent
electric field variance given by \cite{Wang1972:PRB}
\begin{equation}\label{eq:opph}
 \langle E_{z,i}^2\rangle(T) =  \beta_i\frac{\hbar\omega_{s}^{(i)}}{4\pi\varepsilon_0}
\frac{(1+2 n_s^{(i)})}{4 d^3},
\end{equation}
with $\varepsilon_0$ denoting the dielectric constant and
$n_s^{(i)}$ standing for the Bose-Einstein occupation factors of the
phonon mode $i$. The individual strengths of these remote phonon
scattering modes are given by the parameters $\beta_1 = 0.025$ and
$\beta_2 = 0.062$, which fulfill the relation $\beta= \sum_i\beta_i
= (\varepsilon_s-\varepsilon_\infty) /(\varepsilon_s
+1)(\varepsilon_\infty+1)$ with $\beta$ giving a measure of the
total polarizibility of the dielectric interface
\cite{Fratini2008:PRB} and $\varepsilon_s$ and $\varepsilon_\infty$
denoting the static and high-frequency dielectric constant,
respectively.
Due to the randomness of the electrons' motion the spin experiences
an effective electric field and, hence, a random BR-field. The
effective spectral correlation function of the phonon field $\langle
E_z(t) E_z(t')\rangle$ will include an exponential decay with the
momentum relaxation time $\tau_m$ yielding a Lorentzian
renormalization factor $E_{\mathrm{eff},i}^2 = \langle E_{z,i}^2
\rangle/[1+(\omega_{s}^{(i)}\tau_m)^2]$ \cite{Fabian2007:APS}. If
$\omega_s\tau_m \gg 1$ (as for graphene on SiO$_2$), the effective
electric field can be found by qualitative arguments: for long-wave
phonons the spin precesses by $\delta\varphi = \Omega_{\mathrm{BR}}
\tau_\mathrm{ph}$ in the characteristic time $\tau_\mathrm{ph} =
1/\omega_s$. The momentum scattering leads to a random walk with
typical step times of $\tau_m$. The spin is relaxed if the variance
$\langle\delta\varphi\rangle^2 = (t/\tau_m)\delta\varphi$ reaches
one, yielding for the spin relaxation time $1/\tau_s =
\Omega_{\mathrm{BR}}^2 \tau_m /(\omega_\mathrm{ph}\tau_m)^2 =
\Omega_\mathrm{eff}^2\tau_m$ giving an effective field of
$E_\mathrm{eff}^2 = E^2/(\omega_\mathrm{ph}\tau_m)^2$.

\begin{figure}
\includegraphics[width=0.8\linewidth]{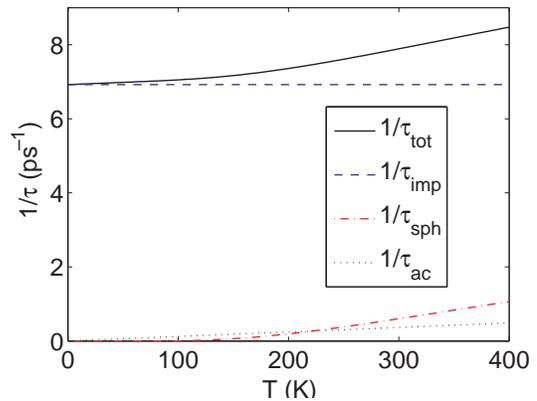}
\caption{\label{fig:taupT} (Color online) Calculated inverse momentum relaxation
times $1/\tau$ as function of temperature $T$ for impurity (imp)
scattering (with $n_{\mathrm{imp}} = 4\times10^{11}$ cm$^{-2}$),
surface phonon (sph), and acoustic phonon (ac) scattering at $E_f =
100$ meV.}
\end{figure}

\begin{figure}
\includegraphics[width=0.9\linewidth]{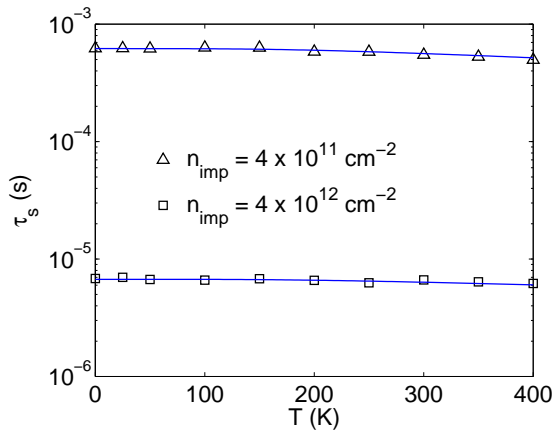}
\caption{\label{fig:tausT} (Color online) Calculated spin relaxation time $\tau_s$
versus temperature $T$ taking into account impurity, surface phonon
and acoustic phonon scattering at $E_f = 100$ meV. The symbols refer
to MC-data and the solid lines are fits of the form $1/\tau_s
=\tau_{\mathrm{tot}}(T) [2 \zeta_{\mathrm{BR}}
E_{\mathrm{eff}}/\hbar]^2$ with $E_{\mathrm{eff}} = 0.21$ V/nm (for
squares) and $E_{\mathrm{eff}} = 0.007$ V/nm (for triangles).}
\end{figure}

For the MC-simulations we took into account momentum scattering due
to charged impurities in SiO$_2$, optical surface phonons
\cite{Fratini2008:PRB}, and acoustic phonons of the graphene sheet
\cite{Chen2008:NatureN}. The resulting total momentum relaxation
rate $1/\tau_{\mathrm{tot}} = 1/\tau_{\mathrm{imp}} +
1/\tau_{\mathrm{sph}}(T)+ 1/\tau_{\mathrm{ac}}(T)$ is illustrated in
Fig.~\ref{fig:taupT}, showing that the impurity scattering remains
dominant up to room temperature but with an exponentially increasing
contribution coming from the surface phonons and a linearly growing
contribution due to acoustic phonon scattering. The random BR-field
is calculated from the electric field originating from the
impurities and the polar surface phonons. The temperature dependence
of the spin relaxation time for a fixed Fermi energy of $E_f = 100$
meV for different impurity densities is shown in Figure
\ref{fig:tausT}, where the solid lines indicate again fits of the
form $1/\tau_s =\tau_{\mathrm{tot}}(T) [2 \zeta_{\mathrm{BR}}
E_{\mathrm{eff}}/\hbar]^2$. The MC-simulations reveal that the spin
relaxation time is almost temperature-independent. This is caused by
the nearly perfect counterbalancing of the increasing electric field
and the decreasing momentum relaxation time with temperature. As for
the relaxation of the momentum, impurities dominate the spin
relaxation compared to the mechanism of optical surface phonons,
which causes a decrease of $\tau_s$ by about $10-20 \%$.

Can we relate our results to the experimental findings of
$\tau_s$ of 100-200 ps \cite{Tombros2007:Nature}? Even considering
the uncertainties in $d$ or in the charge density in the substrate,
such small values for $\tau_s$ can be hardly explained by the substrate
effects.
Indeed, the measured samples have short mean free times of about
$\tau_\mathrm{mfp}\approx 50$ fs \cite{Tombros2007:Nature}, which
suggest a high impurity density of about $n_\mathrm{imp}= 2-4\times
10^{12}$ cm$^{-2}$ \cite{Adam2008:SSC}. However, the times  $\tau_s
\approx 100$ ps would require SO constants orders of magnitude
higher than the ones obtained by first-principles calculations
\cite{Gmitra2009:arXiv} used here. In the experimental samples
graphene was additionally coated by an Al$_2$O$_3$ layer to realize
working tunnel barrier contacts. This likely brings metallic
adatoms, which induce a stronger spin-orbit coupling strength, as
has been reported for a full layer of Au atoms in contact with
graphene, in which a several orders of magnitude larger BR-constant
of about 13 meV was found \cite{Varykhalov2008:PRL}; similar large
SO constants were predicted for impurities on graphene
\cite{Castro2009:arXiv}. Suppose an adatom induces a local
spin-orbit splitting of magnitude $\approx 10$ meV. The splitting
spreads a distance $s$ of perhaps a few bond lengths. Let the
average distance between the randomly positioned adatoms be $r$.
Then the DP spin relaxation rate is $1/\tau_s \approx \Omega^2\tau
(s/r)^2$. The rate is reduced from that for a homogeneous splitting
by $(s/r)^2$, which renormalizes $\Omega^2$ due to the finite
effective adatoms area. As a generic example we take $s$ to be two
bond lengths,  $s\approx 3$ $\AA$, and a reasonable distance $r
\approx 10$ nm, we get the spin relaxation time $\tau_s \approx 50$
ps (using $\tau \approx 100$ fs), being of the same order of
magnitude as the measured value\cite{Tombros2007:Nature}. The adatom
mechanism depends strongly on the adatom type and density, making it
experimentally testable.

In summary, we showed that charged impurities and polar optical surface phonons
of the substrate generate a random Bychkov-Rashba SO-field which
leads to an almost temperature-independent spin relaxation in
graphene. The calculated spin relaxation times give the upper
bounds of what one can expect experimentally for a clean graphene
on a substrate. The above calculation also shows that spin injection and spin
transport should be severely limited if metallic electrodes are deposited directly on
graphene.

This work has been supported by the DFG SFB 689 and SPP 1285. We thank E. Sherman for very valuable
and inspiring discussions.


\begin{thebibliography}{27}
\expandafter\ifx\csname natexlab\endcsname\relax\def\natexlab#1{#1}\fi
\expandafter\ifx\csname bibnamefont\endcsname\relax
  \def\bibnamefont#1{#1}\fi
\expandafter\ifx\csname bibfnamefont\endcsname\relax
  \def\bibfnamefont#1{#1}\fi
\expandafter\ifx\csname citenamefont\endcsname\relax
  \def\citenamefont#1{#1}\fi
\expandafter\ifx\csname url\endcsname\relax
  \def\url#1{\texttt{#1}}\fi
\expandafter\ifx\csname urlprefix\endcsname\relax\def\urlprefix{URL }\fi
\providecommand{\bibinfo}[2]{#2}
\providecommand{\eprint}[2][]{\url{#2}}

\bibitem[{\citenamefont{Geim and Novoselov}(2007)}]{Geim2007:NatureMat}
\bibinfo{author}{\bibfnamefont{A.~K.} \bibnamefont{Geim}} \bibnamefont{and}
  \bibinfo{author}{\bibfnamefont{K.~S.} \bibnamefont{Novoselov}},
  \bibinfo{journal}{Nature Mater.} \textbf{\bibinfo{volume}{6}},
  \bibinfo{pages}{183} (\bibinfo{year}{2007}).

\bibitem[{\citenamefont{Tombros et~al.}(2007)\citenamefont{Tombros, Jozsa,
  Popinciuc, Jonkman, and van Wees}}]{Tombros2007:Nature}
\bibinfo{author}{\bibfnamefont{N.}~\bibnamefont{Tombros}},
  \bibinfo{author}{\bibfnamefont{C.}~\bibnamefont{Jozsa}},
  \bibinfo{author}{\bibfnamefont{M.}~\bibnamefont{Popinciuc}},
  \bibinfo{author}{\bibfnamefont{H.~T.} \bibnamefont{Jonkman}},
  \bibnamefont{and} \bibinfo{author}{\bibfnamefont{B.~J.} \bibnamefont{van
  Wees}}, \bibinfo{journal}{Nature} \textbf{\bibinfo{volume}{448}},
  \bibinfo{pages}{571} (\bibinfo{year}{2007}).

\bibitem[{\citenamefont{Tombros et~al.}(2008)\citenamefont{Tombros, Tanabe,
  Veligura, Jozsa, Popinciuc, Jonkman, and van Wees}}]{Tombros2008:PRL}
\bibinfo{author}{\bibfnamefont{N.}~\bibnamefont{Tombros}},
  \bibinfo{author}{\bibfnamefont{S.}~\bibnamefont{Tanabe}},
  \bibinfo{author}{\bibfnamefont{A.}~\bibnamefont{Veligura}},
  \bibinfo{author}{\bibfnamefont{C.}~\bibnamefont{Jozsa}},
  \bibinfo{author}{\bibfnamefont{M.}~\bibnamefont{Popinciuc}},
  \bibinfo{author}{\bibfnamefont{H.~T.} \bibnamefont{Jonkman}},
  \bibnamefont{and} \bibinfo{author}{\bibfnamefont{B.~J.} \bibnamefont{van
  Wees}}, \bibinfo{journal}{Phys. Rev. Lett.} \textbf{\bibinfo{volume}{101}},
  \bibinfo{pages}{046601} (\bibinfo{year}{2008}).

\bibitem[{\citenamefont{J{\'o}zsa et~al.}(2008)\citenamefont{J{\'o}zsa,
  Popinciuc, Tombros, Jonkman, and van Wees}}]{Jozsa2008:PRL}
\bibinfo{author}{\bibfnamefont{C.}~\bibnamefont{J{\'o}zsa}},
  \bibinfo{author}{\bibfnamefont{M.}~\bibnamefont{Popinciuc}},
  \bibinfo{author}{\bibfnamefont{N.}~\bibnamefont{Tombros}},
  \bibinfo{author}{\bibfnamefont{H.~T.} \bibnamefont{Jonkman}},
  \bibnamefont{and} \bibinfo{author}{\bibfnamefont{B.~J.} \bibnamefont{van
  Wees}}, \bibinfo{journal}{Phys. Rev. Lett.} \textbf{\bibinfo{volume}{100}},
  \bibinfo{pages}{236603} (\bibinfo{year}{2008}).

\bibitem[{\citenamefont{Chen et~al.}(2008{\natexlab{a}})\citenamefont{Chen,
  Jang, Adam, Fuhrer, Williams, and Ishigami}}]{Chen2008:NatureP}
\bibinfo{author}{\bibfnamefont{J.-H.} \bibnamefont{Chen}},
  \bibinfo{author}{\bibfnamefont{C.}~\bibnamefont{Jang}},
  \bibinfo{author}{\bibfnamefont{S.}~\bibnamefont{Adam}},
  \bibinfo{author}{\bibfnamefont{M.~S.} \bibnamefont{Fuhrer}},
  \bibinfo{author}{\bibfnamefont{E.~D.} \bibnamefont{Williams}},
  \bibnamefont{and} \bibinfo{author}{\bibfnamefont{M.}~\bibnamefont{Ishigami}},
  \bibinfo{journal}{Nature Physics} \textbf{\bibinfo{volume}{4}},
  \bibinfo{pages}{377} (\bibinfo{year}{2008}{\natexlab{a}}).

\bibitem[{\citenamefont{Adam and {Das Sarma}}(2008)}]{Adam2008:SSC}
\bibinfo{author}{\bibfnamefont{S.}~\bibnamefont{Adam}} \bibnamefont{and}
  \bibinfo{author}{\bibfnamefont{S.}~\bibnamefont{{Das Sarma}}},
  \bibinfo{journal}{Solid State Commun.} \textbf{\bibinfo{volume}{146}},
  \bibinfo{pages}{356} (\bibinfo{year}{2008}).

\bibitem[{\citenamefont{Sabio et~al.}(2008)\citenamefont{Sabio, Seo\'anez,
  Fratini, Guinea, {Castro Neto}, and Sols}}]{Sabio2008:PRB}
\bibinfo{author}{\bibfnamefont{J.}~\bibnamefont{Sabio}},
  \bibinfo{author}{\bibfnamefont{C.}~\bibnamefont{Seo\'anez}},
  \bibinfo{author}{\bibfnamefont{S.}~\bibnamefont{Fratini}},
  \bibinfo{author}{\bibfnamefont{F.}~\bibnamefont{Guinea}},
  \bibinfo{author}{\bibfnamefont{A.~H.} \bibnamefont{{Castro Neto}}},
  \bibnamefont{and} \bibinfo{author}{\bibfnamefont{F.}~\bibnamefont{Sols}},
  \bibinfo{journal}{Phys. Rev. B} \textbf{\bibinfo{volume}{77}},
  \bibinfo{pages}{195409} (\bibinfo{year}{2008}).

\bibitem[{\citenamefont{Adam et~al.}(2007)\citenamefont{Adam, Hwang, Galitski,
  and {Das Sarma}}}]{Adam2007:PNAS}
\bibinfo{author}{\bibfnamefont{S.}~\bibnamefont{Adam}},
  \bibinfo{author}{\bibfnamefont{E.~H.} \bibnamefont{Hwang}},
  \bibinfo{author}{\bibfnamefont{V.~M.} \bibnamefont{Galitski}},
  \bibnamefont{and} \bibinfo{author}{\bibfnamefont{S.}~\bibnamefont{{Das
  Sarma}}}, \bibinfo{journal}{Proc Natl. Acad. Sci. U S A.}
  \textbf{\bibinfo{volume}{104}}, \bibinfo{pages}{18392}
  (\bibinfo{year}{2007}).

\bibitem[{\citenamefont{Fratini and Guinea}(2008)}]{Fratini2008:PRB}
\bibinfo{author}{\bibfnamefont{S.}~\bibnamefont{Fratini}} \bibnamefont{and}
  \bibinfo{author}{\bibfnamefont{F.}~\bibnamefont{Guinea}},
  \bibinfo{journal}{Phys. Rev. B} \textbf{\bibinfo{volume}{77}},
  \bibinfo{pages}{195415} (\bibinfo{year}{2008}).

\bibitem[{\citenamefont{Guinea}(2008)}]{Guinea2008:JLTP}
\bibinfo{author}{\bibfnamefont{F.}~\bibnamefont{Guinea}}, \bibinfo{journal}{J.
  Low Temp. Phys.} \textbf{\bibinfo{volume}{153}}, \bibinfo{pages}{359}
  (\bibinfo{year}{2008}).

\bibitem[{\citenamefont{Chen et~al.}(2008{\natexlab{b}})\citenamefont{Chen,
  Jang, Xiao, Ishigami, and Fuhrer}}]{Chen2008:NatureN}
\bibinfo{author}{\bibfnamefont{J.}~\bibnamefont{Chen}},
  \bibinfo{author}{\bibfnamefont{C.}~\bibnamefont{Jang}},
  \bibinfo{author}{\bibfnamefont{S.}~\bibnamefont{Xiao}},
  \bibinfo{author}{\bibfnamefont{M.}~\bibnamefont{Ishigami}}, \bibnamefont{and}
  \bibinfo{author}{\bibfnamefont{M.~S.} \bibnamefont{Fuhrer}},
  \bibinfo{journal}{Nature Nanotechnology} \textbf{\bibinfo{volume}{3}},
  \bibinfo{pages}{206} (\bibinfo{year}{2008}{\natexlab{b}}).

\bibitem[{\citenamefont{D'yakonov and Perel'}(1971)}]{Dyakonov1972:SPSS}
\bibinfo{author}{\bibfnamefont{M.~I.} \bibnamefont{D'yakonov}}
  \bibnamefont{and} \bibinfo{author}{\bibfnamefont{V.~I.}
  \bibnamefont{Perel'}}, \bibinfo{journal}{Fiz. Tverd. Tela}
  \textbf{\bibinfo{volume}{13}}, \bibinfo{pages}{3581} (\bibinfo{year}{1971}),
  \bibinfo{note}{[Sov. Phys. Solid State {\bf 13}, 3023-3026 (1971)]}.

\bibitem[{\citenamefont{D'yakonov and Perel'}(1984)}]{Dyakonov:1984b}
\bibinfo{author}{\bibfnamefont{M.~I.} \bibnamefont{D'yakonov}}
  \bibnamefont{and} \bibinfo{author}{\bibfnamefont{V.~I.}
  \bibnamefont{Perel'}}, in \emph{\bibinfo{booktitle}{Optical Orientation,
  Modern Problems in Condensed Matter Science, Vol. 8}}, edited by
  \bibinfo{editor}{\bibfnamefont{F.}~\bibnamefont{Meier}} \bibnamefont{and}
  \bibinfo{editor}{\bibfnamefont{B.~P.} \bibnamefont{Zakharchenya}}
  (\bibinfo{publisher}{North-Holland, Amsterdam}, \bibinfo{year}{1984}),
  p.~\bibinfo{pages}{40}.

\bibitem[{\citenamefont{Fabian et~al.}(2007)\citenamefont{Fabian,
  Matos-Abiague, Ertler, Stano, and {I. \v{Z}uti{\'c}}}}]{Fabian2007:APS}
\bibinfo{author}{\bibfnamefont{J.}~\bibnamefont{Fabian}},
  \bibinfo{author}{\bibfnamefont{A.}~\bibnamefont{Matos-Abiague}},
  \bibinfo{author}{\bibfnamefont{C.}~\bibnamefont{Ertler}},
  \bibinfo{author}{\bibfnamefont{P.}~\bibnamefont{Stano}}, \bibnamefont{and}
  \bibinfo{author}{\bibnamefont{{I. \v{Z}uti{\'c}}}}, \bibinfo{journal}{Acta
  Phys. Slov.} \textbf{\bibinfo{volume}{57}}, \bibinfo{pages}{565}
  (\bibinfo{year}{2007}).

\bibitem[{\citenamefont{Huertas-Hernando
  et~al.}(2007)\citenamefont{Huertas-Hernando, Guinea, and
  Brataas}}]{Huertas2007:EPJ}
\bibinfo{author}{\bibfnamefont{D.}~\bibnamefont{Huertas-Hernando}},
  \bibinfo{author}{\bibfnamefont{F.}~\bibnamefont{Guinea}}, \bibnamefont{and}
  \bibinfo{author}{\bibfnamefont{A.}~\bibnamefont{Brataas}},
  \bibinfo{journal}{Eur. Phys. J. Special Topics}
  \textbf{\bibinfo{volume}{148}}, \bibinfo{pages}{177} (\bibinfo{year}{2007}).

\bibitem[{\citenamefont{Huertas-Hernando
  et~al.}(2008)\citenamefont{Huertas-Hernando, Guinea, and
  Brataas}}]{Huertas2008:condmat}
\bibinfo{author}{\bibfnamefont{D.}~\bibnamefont{Huertas-Hernando}},
  \bibinfo{author}{\bibfnamefont{F.}~\bibnamefont{Guinea}}, \bibnamefont{and}
  \bibinfo{author}{\bibfnamefont{A.}~\bibnamefont{Brataas}}
  (\bibinfo{year}{2008}), \eprint{arXiv:0812.1921v1}.

\bibitem[{\citenamefont{DiVincenzo and Mele}(1984)}]{DiVincenzo1984:PRB}
\bibinfo{author}{\bibfnamefont{D.~P.} \bibnamefont{DiVincenzo}}
  \bibnamefont{and} \bibinfo{author}{\bibfnamefont{E.~J.} \bibnamefont{Mele}},
  \bibinfo{journal}{Phys. Rev. B} \textbf{\bibinfo{volume}{29}},
  \bibinfo{pages}{1685} (\bibinfo{year}{1984}).

\bibitem[{\citenamefont{Kane and Mele}(2005)}]{Kane2005:PRL}
\bibinfo{author}{\bibfnamefont{C.~L.} \bibnamefont{Kane}} \bibnamefont{and}
  \bibinfo{author}{\bibfnamefont{E.~J.} \bibnamefont{Mele}},
  \bibinfo{journal}{Phys. Rev. Lett.} \textbf{\bibinfo{volume}{95}},
  \bibinfo{pages}{226801} (\bibinfo{year}{2005}).

\bibitem[{\citenamefont{Min et~al.}(2006)\citenamefont{Min, Hill, Sinitsyn,
  Sahu, Kleinman, and MacDonald}}]{Min2006:PRB}
\bibinfo{author}{\bibfnamefont{H.}~\bibnamefont{Min}},
  \bibinfo{author}{\bibfnamefont{J.~E.} \bibnamefont{Hill}},
  \bibinfo{author}{\bibfnamefont{N.~A.} \bibnamefont{Sinitsyn}},
  \bibinfo{author}{\bibfnamefont{B.~R.} \bibnamefont{Sahu}},
  \bibinfo{author}{\bibfnamefont{L.}~\bibnamefont{Kleinman}}, \bibnamefont{and}
  \bibinfo{author}{\bibfnamefont{A.~H.} \bibnamefont{MacDonald}},
  \bibinfo{journal}{Phys. Rev. B} \textbf{\bibinfo{volume}{74}},
  \bibinfo{pages}{165310} (\bibinfo{year}{2006}).

\bibitem[{\citenamefont{Yao et~al.}(2007)\citenamefont{Yao, Ye, Qi, Zhang, and
  Fang}}]{Yao2007:PRB}
\bibinfo{author}{\bibfnamefont{Y.}~\bibnamefont{Yao}},
  \bibinfo{author}{\bibfnamefont{F.}~\bibnamefont{Ye}},
  \bibinfo{author}{\bibfnamefont{X.-L.} \bibnamefont{Qi}},
  \bibinfo{author}{\bibfnamefont{S.-C.} \bibnamefont{Zhang}}, \bibnamefont{and}
  \bibinfo{author}{\bibfnamefont{Z.}~\bibnamefont{Fang}},
  \bibinfo{journal}{Phys. Rev. B} \textbf{\bibinfo{volume}{75}},
  \bibinfo{pages}{041401(R)} (\bibinfo{year}{2007}).

\bibitem[{\citenamefont{Huertas-Hernando
  et~al.}(2006)\citenamefont{Huertas-Hernando, Guinea, and
  Brataas}}]{Huertas2006:PRB}
\bibinfo{author}{\bibfnamefont{D.}~\bibnamefont{Huertas-Hernando}},
  \bibinfo{author}{\bibfnamefont{F.}~\bibnamefont{Guinea}}, \bibnamefont{and}
  \bibinfo{author}{\bibfnamefont{A.}~\bibnamefont{Brataas}},
  \bibinfo{journal}{Phys. Rev. B} \textbf{\bibinfo{volume}{74}},
  \bibinfo{pages}{155426} (\bibinfo{year}{2006}).

\bibitem[{\citenamefont{Gmitra et~al.}(2009)\citenamefont{Gmitra, Konschuh,
  Ertler, Ambrosch-Draxl, and Fabian}}]{Gmitra2009:arXiv}
\bibinfo{author}{\bibfnamefont{M.}~\bibnamefont{Gmitra}},
  \bibinfo{author}{\bibfnamefont{S.}~\bibnamefont{Konschuh}},
  \bibinfo{author}{\bibfnamefont{C.}~\bibnamefont{Ertler}},
  \bibinfo{author}{\bibfnamefont{C.}~\bibnamefont{Ambrosch-Draxl}},
  \bibnamefont{and} \bibinfo{author}{\bibfnamefont{J.}~\bibnamefont{Fabian}}
  (\bibinfo{year}{2009}), \eprint{arXiv:0904.3315}.

\bibitem[{\citenamefont{Rashba}(2009)}]{Rashba2009:PRB}
\bibinfo{author}{\bibfnamefont{E.~I.} \bibnamefont{Rashba}},
  \bibinfo{journal}{Phys. Rev. B} \textbf{\bibinfo{volume}{79}},
  \bibinfo{pages}{161409(R)} (\bibinfo{year}{2009}).

\bibitem[{\citenamefont{Sherman}(2003)}]{Sherman2003:APL}
\bibinfo{author}{\bibfnamefont{E.~Y.} \bibnamefont{Sherman}},
  \bibinfo{journal}{Appl. Phys. Lett.} \textbf{\bibinfo{volume}{82}},
  \bibinfo{pages}{209} (\bibinfo{year}{2003}).

\bibitem[{\citenamefont{Wang and Mahan}(1972)}]{Wang1972:PRB}
\bibinfo{author}{\bibfnamefont{S.~Q.} \bibnamefont{Wang}} \bibnamefont{and}
  \bibinfo{author}{\bibfnamefont{G.~D.} \bibnamefont{Mahan}},
  \bibinfo{journal}{Phys. Rev. B} \textbf{\bibinfo{volume}{6}},
  \bibinfo{pages}{4517} (\bibinfo{year}{1972}).

\bibitem[{\citenamefont{Varykhalov et~al.}(2008)\citenamefont{Varykhalov,
  S{\'a}nchez-Barriga, Shikin, Biswas, Vescovo, Rybkin, Marchenko, and
  Rader}}]{Varykhalov2008:PRL}
\bibinfo{author}{\bibfnamefont{A.}~\bibnamefont{Varykhalov}},
  \bibinfo{author}{\bibfnamefont{J.}~\bibnamefont{S{\'a}nchez-Barriga}},
  \bibinfo{author}{\bibfnamefont{A.~M.} \bibnamefont{Shikin}},
  \bibinfo{author}{\bibfnamefont{C.}~\bibnamefont{Biswas}},
  \bibinfo{author}{\bibfnamefont{E.}~\bibnamefont{Vescovo}},
  \bibinfo{author}{\bibfnamefont{A.}~\bibnamefont{Rybkin}},
  \bibinfo{author}{\bibfnamefont{D.}~\bibnamefont{Marchenko}},
  \bibnamefont{and} \bibinfo{author}{\bibfnamefont{O.}~\bibnamefont{Rader}},
  \bibinfo{journal}{Phy. Rev. Lett.} \textbf{\bibinfo{volume}{101}},
  \bibinfo{pages}{157601} (\bibinfo{year}{2008}).

\bibitem[{\citenamefont{{Castro Neto} and Guinea}(2009)}]{Castro2009:arXiv}
\bibinfo{author}{\bibfnamefont{A.~H.} \bibnamefont{{Castro Neto}}}
  \bibnamefont{and} \bibinfo{author}{\bibfnamefont{F.}~\bibnamefont{Guinea}}
  (\bibinfo{year}{2009}), \eprint{arXiv:0902.3244}.

\end{thebibliography}

\end{document}